\documentclass[superscriptaddress,prd,nofootinbib,preprintnumbers,longbibliography,11pt,eqsecnum]{revtex4-1}

\usepackage{amsmath} 
\usepackage{graphicx} 
\usepackage{amsthm}
\usepackage{amssymb} 
\usepackage{dsfont}
\usepackage{yfonts}
\usepackage{hyperref}
\usepackage{array,xcolor,graphicx}
\usepackage{booktabs,multirow}
\usepackage[utf8]{inputenc}
\usepackage{mathtools}

\usepackage{slashed}
\usepackage{caption}
\usepackage{stackrel}
\usepackage{cases}

\usepackage{etoolbox}
\patchcmd{\section}
  {\centering}
  {\raggedright}
  {}
  {}
\patchcmd{\subsection}
  {\centering}
  {\raggedright}
  {}
  {}

\usepackage[normalem]{ulem}
\usepackage{mathtools}

\usepackage[all]{xy}
\usepackage{tikz}
\usetikzlibrary{arrows.meta}

\hypersetup{colorlinks=true,linkcolor=blue,citecolor=red,urlcolor=blue, linktocpage=true}

\newcommand{\be}{\begin{equation}}
\newcommand{\ee}{\end{equation}}
\newcommand{\bea}{\begin{eqnarray}}
\newcommand{\eea}{\end{eqnarray}}

\def\la{\langle}
\def\ra{\rangle}

\def\d{\partial}

\def\CB{\mathcal{B}}

\def\CJ{\mathcal{J}}

\def\CN{\mathcal{N}}

\def\CO{\mathcal{O}}
\def\CP{\mathcal{P}}

\def\ttr{\mathtt{r}}

\graphicspath{ {./images/} }

\begin{document}
	
	\title{Operator lifetime and the force-free electrodynamic limit \\ 
	of magnetised holographic plasma}
	
	\author{Napat Poovuttikul}
\email{napat.poovuttikul@durham.ac.uk}
\affiliation{Department of Mathematical Sciences,
Durham University, South Road, Durham DH1 3LE, UK}
\affiliation{University of Iceland, Science Institute, Dunhaga 3, IS-107, Reykjavik, Iceland}

	\author{Aruna Rajagopal}
	\email{arr17@hi.is}
\affiliation{University of Iceland, Science Institute, Dunhaga 3, IS-107, Reykjavik, Iceland}

	\begin{abstract}
	\vspace{1cm}
	Using the framework of higher-form global symmetries, we examine the regime of validity of force-free electrodynamics by evaluating the lifetime of the electric field operator, which is non-conserved due to screening effects. We focus on a holographic model which has the same global symmetry as that of low energy plasma and obtain the lifetime of (non-conserved) electric flux in a strong magnetic field regime. The lifetime is inversely correlated to the magnetic field strength and thus suppressed in the strong field regime.  
	\end{abstract}
	\maketitle
	\begingroup
	\hypersetup{linkcolor=black}
	\tableofcontents
	\endgroup
	\hrulefill

\section{Introduction}

\begin{figure}[tbh]
\centering
\includegraphics[width=0.5\textwidth]{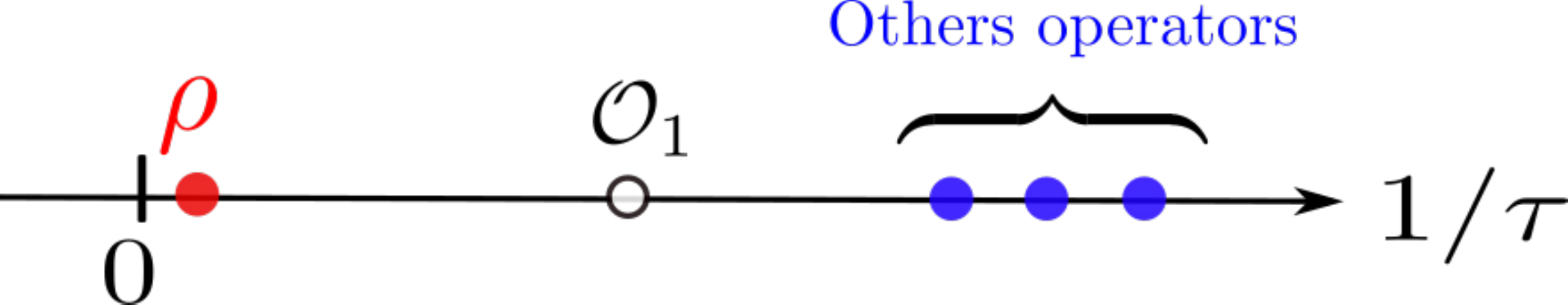}
    \captionsetup{justification=raggedright,singlelinecheck=false}
\caption{ A cartoon illustration of the lifetime of operators of a theory that exhibit hydrodynamic behaviour at late time. Here, there is a parametrically large gap between conserved charges $\rho$ and the rest. The life time $\tau_1$ of the longest-lived operator, denoted by $\CO_1$ set the time scale in which hydrodynamics becomes applicable.  
}
\label{fig:lifetime}
\end{figure}

Hydrodynamics \cite{LLfluid} is a well-established theoretical framework which universally describes the long wavelength, low frequency behaviour of interacting systems at finite temperature.
Essentially, hydrodynamic theory is a description of conserved quantities and the manifestation of the corresponding symmetries in a system in thermal equilibrium. 
Theories with widely varying microscopics can have the same macroscopic hydrodynamic description. One possible explanation why such a universal description is possible is that all operators except conserved charges have parametrically short lifetimes compared to the scale of interest and, once the longest-lived non-conserved operator\footnote{While this operator language is more familiar in the context of quantum systems, it is also applicable to classical systems via e.g. memory matrix formalism \cite{zwanzig1995,forster1995}. A more modern introduction may be found in \cite{Hartnoll:2012rj}.} has decayed away, the hydrodynamic description becomes viable (see Fig. \ref{fig:lifetime}). 

        The hydrodynamic framework may be generalised to systems where the conserved charges are those of a higher-form symmetry \cite{Gaiotto:2014kfa} which counts the number density of extended objects. A recent exploration of this idea \cite{Grozdanov_2017} (see also \cite{Schubring:2014iwa,Hernandez:2017mch,Armas:2018atq,Armas:2018zbe}) shows that the resulting hydrodynamics of a one-form $U(1)$ charge reproduces the theory of magnetohydrodynamics (MHD)\footnote{For the formulation of MHD that closely resembles higher-form symmetry formulation, see e.g. \cite{dixon1982special,anile2005relativistic,komissarov1999}.}. This should not come as a big surprise. MHD is, after all, a low energy effective theory of plasma where the (dynamical) electric field is screened -- the one-form U(1) symmetry associated to electric flux is explicitly broken. This implies that in, for example, a plasma at zero magnetic field (where the Ohm's law ${\bf j} = \sigma {\bf E}$ is a good approximation) the electric field has a finite lifetime,
\begin{equation}\label{eq:lifetimeGen}
    \delta {\bf E} \propto \exp(-t/\tau_E ) \qquad \Longleftrightarrow\qquad  \la E^i(-\omega) E^j(\omega)\ra \sim \frac{\delta^{ij}}{\omega + i/\tau_E}     \,, 
\end{equation}
with the lifetime of the electric field $\tau_E = 1/\sigma$. The conductivity $\sigma$ can be computed from first principles. For instance, in quantum electrodynamics, it can be written as \cite{Arnold_2000}\footnote{The fact that this quantity has only been computed at the beginning of this century indicates the difficulty of the required computations.  }
\begin{equation}\label{eq:lifetimeZeroB}
    \sigma \propto \frac{T}{e^2 \log e^{-1}}\,,
\end{equation}
where $e$ is the electromagnetic coupling. The lifetime of electric field $\tau \sim 1/T$, is then much shorter than the scale $t \gg 1/T$ (or $\omega/T \ll 1$ in Fourier space) where hydrodynamic behaviour is expected. If, in this late-time limit, all other operators except energy density $T^{tt}$ and the momentum $T^{ti}$ have already decayed away, one can expect the hydrodynamic description of a plasma to be governed by
\begin{equation}\label{eq:MHDWardIden}
    \d_\mu T^{\mu\nu} = 0 \, , \qquad \d_\mu J^{\mu\nu} = 0\,.
\end{equation}
The conserved currents $T^{\mu\nu}$ and $J^{\mu\nu}$ are expressed in terms of energy, momentum, magnetic flux $J^{ti}\equiv B^i$ and their conjugates, organised order by order in the gradient expansion. This formulation of MHD only requires macroscopic consistency and does not require the introduction of the gauge field $\star J = F = dA$ which, due to screening effect, is not a long-lived degree of freedom. 

This brings us to the central question of the present paper: Is a hydrodynamic description of the form \eqref{eq:MHDWardIden} applicable in the limit low temperature compared to magnetic flux density $T^2/|{\bf B}| \ll1$ ? This question is important if one wants to apply the MHD description to astrophysical plasmas where the magnetic field is many orders of magnitude larger than the scale set by the temperature. 
%
%
If one were to naively extrapolate \eqref{eq:lifetimeGen}-\eqref{eq:lifetimeZeroB}, the lifetime of the electric field appears to become arbitrarily long as the temperature decreases. However, there exists a macroscopic description of plasma in this regime that has been successfully applied. This theory is called \textit{force-free electrodynamics} or FFE, and has been used extensively in astrophysical setups such as in the magnetosphere of black holes \cite{Blandford:1977ds,Komissarov_2004}, neutron star \cite{1969ApJ...157..869G} and solar corona \cite{Wiegelmann_2012} just to name a few. In its conventional form, this theory is applied to a system which is magnetically dominated (i.e. $|{\bf B}^2| > |{\bf E}^2|$ or, covariantly $F_{\mu\nu}F^{\mu\nu} > 0$) and whose dynamics is governed by 
\begin{subequations}\label{eq:FFEconventionalForm}
\begin{align}
    \epsilon^{\mu\nu\rho\sigma} F_{\mu\nu}F_{\rho\sigma} &= 0 \,,\label{eq:FFEconventionalForm-1}\\
    F^{\mu\nu} \nabla_\lambda F^{\lambda}_{\;\;\nu} &=0 \,. \label{eq:FFEconventionalForm-2}
\end{align}
\end{subequations}
Here, the first relation implies that ${\bf E}\cdot {\bf B} =0$ while the second relation implies that the \textit{force} $j_{el}^\mu F_{\mu\nu}$, with $j^\mu_{el} := \nabla_\nu F^{\nu\mu}$ via Maxwell's equations, acting on plasma vanishes (hence the name force-free electrodynamics). More details on the geometric and effective action view point of FFE can be found in e.g. \cite{Gralla_2014,Compere:2016xwa} and \cite{Uchida:1997,Thompson:1998ss,Gralla_2019,glorioso2018effective}. One should emphasise that the system of equations in Eq.\eqref{eq:FFEconventionalForm} is independent of the microscopic details of the cold plasma, which then strongly resembles hydrodynamic descriptions. In fact, it turns out that \eqref{eq:FFEconventionalForm} can arise in a special limit where $T \ll \sqrt{|{\bf B}|}$ of a hydrodynamics description with one-form $U(1)$ symmetry in \eqref{eq:MHDWardIden}, see \cite{Grozdanov_2017,Gralla_2019,glorioso2018effective}\footnote{Recasting of force-free electrodynamics in the hydrodynamic language also allows the systematic gradient expansions \cite{Gralla_2019,Benenowski:2019ule}. This could serve to classify correction to FFE in order to account for phenomena such as pulsar radio emission where ${\bf E}\cdot {\bf B}\ne 0$. }. 

The existence of FFE is usually justified by saying that the cold plasma is, on one hand, dense enough to screen the electric field \eqref{eq:FFEconventionalForm-1} but, on the other hand, dilute enough so that force-free condition \eqref{eq:FFEconventionalForm-2} is applicable. This statement can be made more precise in the light of relations between the equations of FFE and hydrodynamics. Thus, we propose a criterion for testing the validity of FFE using the lifetime of non-conserved operators -- \textit{FFE, or equivalently, hydrodynamic description of cold plasma in the $T \ll \sqrt{|{\bf B}|}$ limit, is valid when the lifetime of all non-conserved operators is parametrically shorter than the time scale of interest.} A key advantage of this approach is that the operator lifetime can be, in principle, computed explicitly from microscopic description and therefore allows one to find the `cutoff' scale where FFE description should break down. 

Computing the operator lifetime from microscopic description is, however, not always an easy task. In fact, we are not aware of a genuine computation directly from quantum electrodynamics (in the sense of \cite{Arnold_2000}) when both $T$ and ${\bf B}$ are turned on. To simplify the computations, we shall demonstrate the validity of FFE in the strongly interacting magnetised plasma with a holographic dual as proposed in \cite{Grozdanov_2019b,Hofman_2018} where the one-form $U(1)$ global symmetry is taken into account via a  two-form gauge field in the gravity dual. This provides two key advantages. First, the computation of correlation functions boils down to solving simple linearised differential equations (see e.g. \cite{Son:2002sd}). Second, there is strong evidences that charge neutral operators, apart from energy and momentum, have a parametrically short lifetime in this class of theories \footnote{To be more precise, it has been shown in $\CN=4$ supersymmetric Yang-Mills theory, which constitutes the matter sector of the holographic model \cite{Grozdanov_2019b,Hofman_2018}, that there is no long-lived mode besides hydrodynamic modes at any $T\ne 0$ and $|{\bf B}|=0$ \cite{Kovtun:2003wp}. A similar conclusion was reached for the same theory in the charge neutral sector at finite \textit{non-dynamical} magnetic field \cite{Fuini:2015hba,Janiszewski:2015ura}.  }. Therefore, we shall focus on non-conserved operators in the electromagnetic sector of the theory: the electric flux operators, whose lifetime can be extracted via two-point correlation function as in \eqref{eq:lifetimeGen}. This will provide strong evidence for the validity of FFE limit in a strongly interacting holographic plasma. 

On the technical side, the computations presented in this note  show that there are no quasinormal modes present in the vicinity of the hydrodynamic regime $\omega/T\ll 1$ (and $\omega/\sqrt{|{\bf B}|}\ll 1$). The pole in the electric flux correlation function in this regime then implies that the operator has a parametrically long lifetime which could interfere with the hydrodynamic modes. The presence of such long-lived mode can be determined analytically in the usual hydrodynamic regime of $\omega/T \ll 1$ for a large class of theories. It is usually difficult to go beyond this regime towards the limit $\omega/T\sim 1, \omega/\sqrt{|{\bf B}|}\ll 1$. Such computation can, however, be done analytically in the simple model of \cite{Grozdanov_2019b} thanks to the presence of the BTZ$\times \mathbb{R}^2$ bulk geometry in the deep IR \cite{DHoker:2009mmn}. We should also note that the treatment of a (long-lived) non-hydrodynamic modes has been extensively used to determine the breakdown of hydrodynamic descriptions in the context of QFTs with holographic duals, see e.g. \cite{Grozdanov_2019,Davison:2014lua,Chen:2017dsy,Davison:2018nxm}
.

The remainder of this paper is organised as the follows. In section \ref{sec:holomodel}, we summarise the procedure involved in the computation of the two-point correlation function in the holographic dual to one-form global symmetry. In section \ref{sec:matchingwT}, we outline the method for exploring the existence of decaying modes in the vicinity of the usual hydrodynamic limit $\omega/T \ll 1$ at $T/\sqrt{{\bf B}|} \ll 1$. Due to the simplicity of the bulk geometry, we are able to further extend the analysis to arbitrary value of $\omega/T$ with $\omega/\sqrt{|{\bf B}|} \ll 1$ and $T/\sqrt{|{\bf B}|} \ll1$. This is described in section \ref{sec:matchingwB}. Further open questions and future directions are discussed in section \ref{sec:discuss}. 



\section{The Holographic Model}\label{sec:holomodel}

A simple holographic dual to a strongly interacting field theory of matter charged under dynamical $U(1)$ electromagnetism (that is, the dynamical plasma described by low energy MHD) and formulated in the language of higher-form symmetry was constructed in \cite{Grozdanov_2019b,Hofman_2018}. We present a brief review here for completeness. The five-dimensional bulk theory is comprised of Einstein gravity coupled to a two-form bulk gauge field, $B_{\mu\nu}$, and a negative cosmological constant, 

\begin{equation}
  S =  \int d^5X \sqrt{-G} \left( R- 2\Lambda - \frac{L^2}{3} H_{abc}H^{abc} \right)+ S_{bnd} - \frac{1}{\kappa(\Lambda)}\int_{r=\Lambda} d^4x\sqrt{-\gamma}(n^aH_{a\mu\nu})(n_bH^{b\mu\nu}),
\end{equation}
where $H=dB$ and $B_{ab}$ is the bulk 2-form gauge field, $\Lambda$ is the UV-cutoff, $n^a$ is the unit normal to the boundary, and $S_{bnd}$ denotes the Gibbons-Hawking and gravitational counter term. Roughly speaking, the two bulk fields $G_{ab}$ and $B_{ab}$, asymptote to $g_{\mu\nu}$ and $b_{\mu\nu}$ respectively, which then source the currents, $T^{\mu\nu}$ and $J^{\mu\nu}$.
\begin{equation}
    \la T_{\mu\nu}\ra  \equiv \frac{2}{\sqrt{-g}}\frac{\delta S}{\delta g_{\mu\nu}}\,, \qquad  \la J_{\mu\nu}\ra \equiv \frac{1}{\sqrt{-g}}\frac{\delta S}{\delta b_{\mu\nu}}
\end{equation}
The generating functional takes the form,
\begin{equation}
    Z[g_{\mu\nu},b_{\mu\nu}] = \left\langle \exp\left[ i\int d^4x \sqrt{-g}\left(\frac{1}{2}T^{\mu\nu} g_{\mu\nu} + J^{\mu\nu}b_{\mu\nu}  \right)  \right]\right\rangle
\end{equation}
and diffeomorphism invariance and gauge symmetry lead to the following equations,
\begin{equation}
    \nabla_{\mu}\la T^{\mu\nu}\ra  = (db)^{\nu}_{\rho\sigma} \la J^{\rho\sigma}\ra \,,\qquad \nabla_{\mu} \la J^{\mu\nu}\ra = 0.
\end{equation}
$H = db$ is the three-form field strength of the two-form external source. The equilibrium solution of this holographic model is a domain wall interpolating between an asymptotic $AdS_5$ geometry in the UV ($r\to\infty$ in our convention), and $BTZ \times \mathbb{R}^2$ in the near-horizon IR ($r=r_h$). It is described by the following metric and gauge field
\begin{equation}
\begin{aligned}
  ds^2 &= G_{ab} dX^a dX^b = -r^2f(r) dt^2 + \frac{dr^2}{r^2 f(r)} + e^{2V(r)}(dx^2 +dy^2) + e^{2W(r)} dz^2\, ,\\
 B &= h(r) dt \wedge dz \qquad \text{with} \qquad \star_5 H= \CB dx\wedge dy
  \end{aligned}
\end{equation}
Modulo the subtleties due to the mixed boundary conditions, this is nothing but the hodge dual of the magnetised black brane solution of \cite{DHoker:2009mmn}. The radial coordinate is chosen such that $r\to \infty$ corresponds to the usual asymptotic $AdS_5$ with
\begin{equation}
    f(r) =1\, ,\qquad e^{2V(r)} = e^{2W(r)} = r^2
\end{equation}
in the $r\to \infty$ limit. The $BTZ\times \mathbb{R}^2$ solution near the horizon can be written as 
\begin{equation}\label{eq:solBGBTZR2}
    f(r) = 3 \left( 1- \frac{r_h^2}{r^2} \right)\, , \qquad e^{2V} = \frac{\CB}{\sqrt{3}}\, ,\qquad e^{2W} = 3r^2\, .
\end{equation}
The temperature is set by the horizon radius via $4\pi T = r_h^2\vert f'(r_h)\vert= 6r_h/L^2$. We set $L=1$ for simplicity. Note also that $\CB$ is related to the $z-$component of the `physical' magnetic field ${\bf B}$ which differs by a prefactor $L$ or the 2-form gauge field coupling in the bulk (e.g. if one were to define the action with $S\sim \int (1/g^2) H^2$). We will keep using $\CB$ to emphasise its holographic origin but there is no harm in thinking of it as simply ${\bf B}$. 

One interesting feature of this model is that the leading divergence of $B_{\mu \nu}$ in the Fefferman-Graham expansion is logarithmic. Thus, the definition of the source $b_{\mu\nu}$ requires mixed boundary condition
\begin{equation}
\label{eq:defSourceAndResponse}
    b_{\mu\nu} = B_{\mu\nu}(\Lambda) - \frac{1}{\kappa(\Lambda)} \la J_{\mu\nu}\ra \,,\qquad \text{with}\qquad \la J^{\mu\nu}\ra =- \sqrt{-G} n_\alpha H^{\alpha \mu\nu}
\end{equation}
Requiring the source $b_{\mu\nu}$ to be independent of the UV cutoff fixes the form of the `coupling constant' $1/\kappa(\Lambda)$ which turns out to be logarithmically running. This is a common feature for fields with this type of near-boundary behaviour where the counterterm also plays the role of the double-trace deformation \cite{Witten:2001ua,Berkooz:2002ug}, see also \cite{Hofman_2018,Grozdanov_2019b}
for a discussion in the present context. Mapping $J^{\mu\nu}$ in to a more familiar \textit{dynamical} field strength via $J^{\mu\nu} = \frac{1}{2}\mathcal{\epsilon}^{\mu\nu\rho\sigma}F_{\rho\sigma}$, one can see that the double-trace deformation plays a role similar to the Maxwell term for the dynamical gauge field in the dual QFT with $1/\kappa(\Lambda)$ as a (logarithmically running) electromagnetic coupling.

The finite part of $1/\kappa(\Lambda)$ plays a crucial role in this setup. While the finite counterterm in the ordinary bulk Maxwell theory simply results in a contact term in the correlation function, the mixed boundary condition for $B_{ab}$ implies the existence of the purely decaying mode $\omega = -i/ \tau_E$ that can interfere with the gapless hydrodynamic excitation. This is nothing but the life-time of the electric flux operator $Q_E \sim \int dS_{ij} J^{ij}$ which appears in the following correlation function \cite{Hofman_2018,Grozdanov_2019}
\begin{equation}
    \la J^{ij}(t) J^{kl}(0)\ra \sim  \exp\left(-i t/\tau_E\right)\, .
\end{equation}
Note that, due to the anisotropy introduced by finite equilibrium magnetic field, the value of $\tau_E$ depends on which direction of the electric field in consideration.
The limit where $\tau_E$ is small, but finite, compared to the length scale of interest (set by temperature or magnetic flux density) is of particular interest as it allows one to extract $\tau_E$ analytically, via a matching procedure that we outline below. As argued in the introduction, the lifetime of the electric flux determines the validity of MHD and FFE description.

\subsection{Linearised solutions in $\omega/T \ll 1$ limit and matching procedure}\label{sec:matchingwT}

In this section, we outline the computation required to obtain the relaxation time of the electric field. We focus on the hydrodynamic regime where $\omega/T\ll 1$, and the low temperature limit
\footnote{Similar computation for the holographic theory dual to a system with ordinary(zero-form) $U(1)$ symmetry can be found in e.g. \cite{Davison:2013bxa,Moitra:2020dal}. 
}
$T/\sqrt{|{\bf B}|} \ll 1$. This allows us to solve the bulk equation of motion analytically via a matching method similar to that was employed in \cite{Kovtun:2003wp} (see also \cite{Grozdanov_2019} for a recent review). We consider the decay rate of the electric field both along and perpendicular to the equilibrium magnetic field denoted by $E^\parallel = J^{xy}$ and $E^\perp = J^{xz},J^{yz}$ respectively. 

Before proceeding, let us summarise the matching procedure for the $\omega/T\ll 1$ expansion. It involves separating the bulk into three suitably defined pieces: inner region, intermediate region and outer region. The inner region is a suitably defined region close to the horizon while the outer region is defined to be the range of $r$ such that $\omega/r \ll 1$ so that one can drop terms quadratic in $(\omega/r)^2$, which includes the near boundary region. The integration constants of the solution in the outer region are determined by matching the form of inner region solution for intermediate value of $r$ that connect the two regions together. In our case, this is the region of $r$ close to $r_h$ but 
\begin{equation}
\frac{\omega}{T} \log f(r) \ll 1
\end{equation}
This intermediate region defined above is also consistent with the outer region assumption where $\omega/r \ll 1$ and thus we are able to match the two solution together. Note that, while this procedure is applicable to any bulk solution with event horizon, the limit $\omega/T \ll 1$ is crucial. 

We now present the key equations and resulting lifetime of the electric flux.

\subsubsection{Perturbation parallel to equilibrium magnetic field}\label{sec:matching-parallel}
As the magnetic field in equilibrium points along the  $z-$direction, we are interested in $E^\parallel = \frac{1}{2} \varepsilon^{zxy} \la J_{xy}\ra$. The corresponding bulk perturbation is $\delta B_{xy}$ which decouples from the metric perturbation in the zero wave vector limit. The bulk equation of motion can be written as 
\begin{equation} \label{eq:BxyBulkEom}
    \left( r^2 f e^{W-2V} \delta B_{xy}'  \right)' + \frac{\omega^2}{r^2f} e^{W-2V} \delta B_{xy} = 0 
\end{equation}
where $(...)'$ denotes a derivative w.r.t. the radial coordinate $r$. The inner region solution for $\delta B_{xy}$, where we substitute the $BTZ\times \mathbb{R}^2$ solution for $f,V,W$, with the ingoing boundary condition can be written as 
\begin{equation}\label{eq:innerBxy}
\delta B_{xy}^{inner} = c^H \exp\left( -\frac{i\omega}{4\pi T} \log f(r) \right)
    \end{equation}
The outer region solution can be obtained by considering the solution at linear order in $\omega/r$ and one obtains,
\begin{equation}\label{eq:outerBxy}
\begin{aligned}
\delta B_{xy}^{outer}(r) &= c_1 - c_2 \left(\log  \Lambda - \int^\Lambda_{\ttr=r} d\ttr \, \frac{e^{2V(\ttr)-W(\ttr)}}{\ttr^2f(\ttr)}\right)  \\
&= c_1 - c_2 \left(\log r - \phi(r)+ \frac{e^{2V-W}}{r_h^2 f'}\Big\vert_{r=r_h} \log  f \right)\,,
    \end{aligned}
    \end{equation}
where $\phi(r)$ is a function regular everywhere in the bulk defined as 
\begin{equation}
\phi(r) = \int^{\Lambda}_{\ttr=r} d\ttr\, \left[ \frac{e^{2V(\ttr)-W(\ttr)}}{\ttr^2f(\ttr)} - \left(\frac{e^{2V(\ttr)-W(\ttr)}}{r_h^2f'(\ttr)}  \right)_{\ttr=r_h} \frac{f'(\ttr)}{f(\ttr)}  - \frac{1}{\ttr} \right] \,.\nonumber
\end{equation}
This parametrisation allows us to single out leading contributions that dominate when considering the solution near $r=\Lambda$, where $\phi(r)$ and $\log (e^{-2V}r^2f)$ vanish, as well as near $r\approx r_h$ where the $\log f$ term dominates. 
The integration constants $c_1,c_2$ in \eqref{eq:outerBxy} are related to the source $b_{\mu\nu}$ and the 2-form current $\la J^{xy}\ra$. The precise relations can be obtained via Eq. \eqref{eq:defSourceAndResponse} to be 
\begin{equation}
    \la J^{xy}\ra = c_2 \, ,\qquad b_{xy} = c_1 - \left(\log\,\Lambda + \frac{1}{\kappa(\Lambda)}  \right) c_2
\end{equation}
%
Note that, for the source to be independent of the UV cutoff, one requires $\kappa(\Lambda)^{-1} = \text{finite term} - \log \Lambda$. This is the logarithmically running coupling usually found in a double-trace deformed theory and resembles the running of electromagnetic coupling as pointed out in \cite{Grozdanov_2019b,Hofman_2018,Grozdanov_2019}.

For the outer and inner region solutions to match, we consider both solutions in the intermediate region where we can write the inner solution as 
\begin{equation}
    \exp\left(-\frac{i\omega}{4\pi T}\log  f \right)\approx 1 - i\frac{\omega}{4\pi T} \log f + \CO\left(\frac{\omega}{T}  \right)^2
    \label{eq:expandInnerToIntermediate}
\end{equation}
The matching condition $\delta B_{xy}^{inner} = \delta B_{xy}^{outer}$ in this region prompts yield the following algebraic relations between the boundary quantities $b_{xy},\la J_{xy}\ra$:
\begin{equation}
\begin{aligned}
    \frac{i\omega}{4\pi T}c^H &= \left( \frac{\CB/r_h}{3r_h^2f'(r_h)} \right) \la J_{xy}\ra \,\\
     c^H &= b_{xy} + \left[ \frac{1}{\kappa(\Lambda)}+\log\left(\frac{\Lambda}{r_h}   \right) + \phi(r_h) \right]\la J_{xy}\ra\,.
\end{aligned}
\end{equation}
Solving these equations at vanishing source $b_{xy}=0$ yields the spectrum of the form $\omega = -i/\tau_{E^\parallel}$ where $\tau_{E^\parallel}$ is the lifetime of the electric flux parallel to the equilibrium magnetic field. This is the first key result that we advertised earlier, namely
\begin{equation}\label{eq:lifetimeEparallel}
    \tau_{E^{\parallel}} = \frac{2\pi T}{\CB} \left( e_r^{-2} + \phi(r_h)\right)\,,
\end{equation}
where we write $e_r^{-2} =  \log(\Lambda/r_h)+ \kappa(\Lambda)^{-1}$ which plays the role of renormalised electromagnetic coupling. More details on the $T/\sqrt{\CB}$ dependence of $\phi(r_h)$ can be found in Appendix \ref{app:numerics}.      

\begin{figure}[tbh]
\centering
\includegraphics[width=0.55\textwidth]{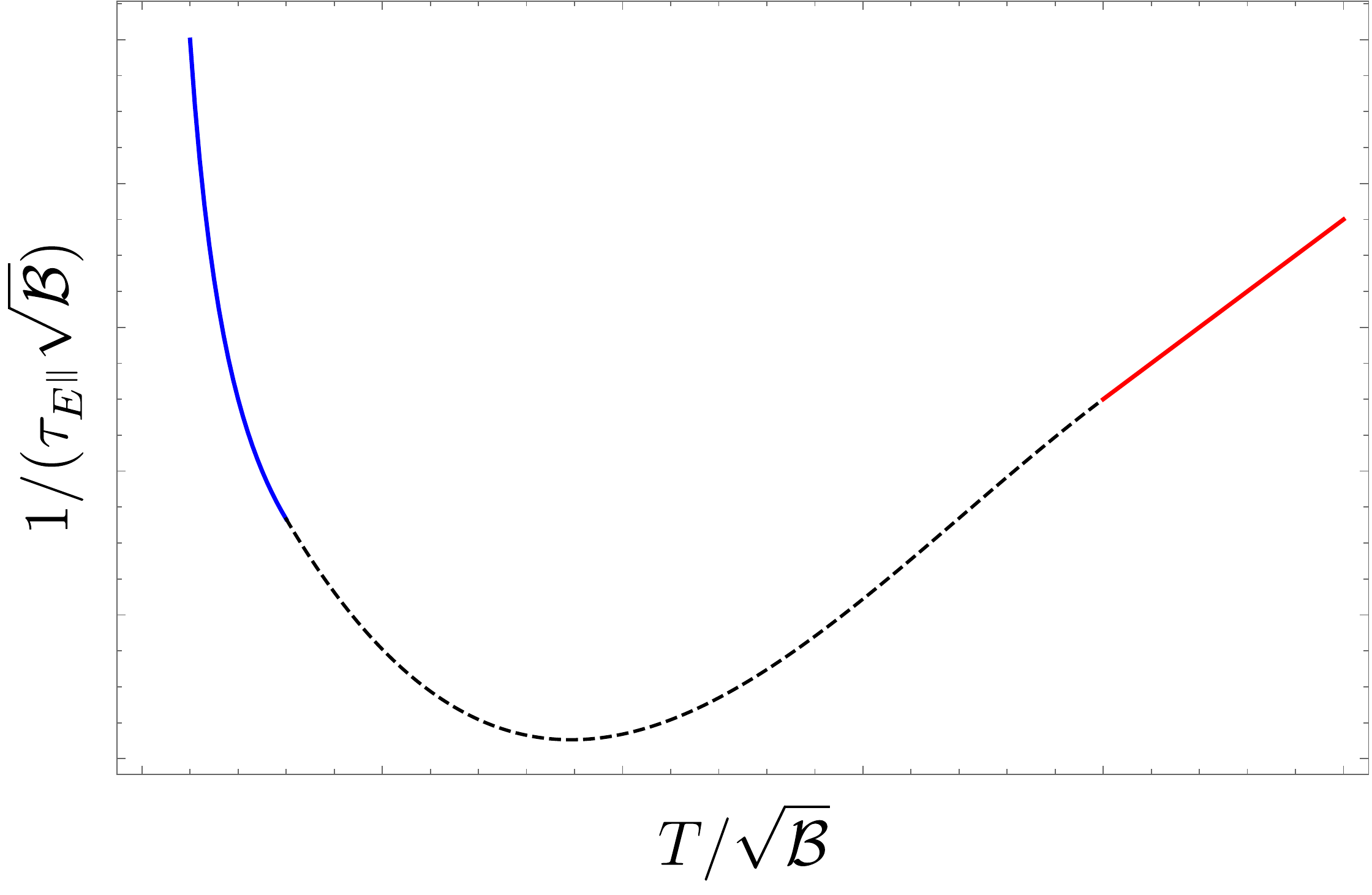}
    \captionsetup{justification=raggedright,singlelinecheck=false}
\caption{ A sketch of the decay rate (inverse of the lifetime) of the electric field as a function of $T/\sqrt{\CB}$, measured in the unit of $\sqrt{\CB}$. The high temperature regime (red) depict the result of decay rate at zero magnetic field found in \cite{Hofman_2018,Grozdanov_2019} which has the same temperature dependence as in \eqref{eq:lifetimeGen}-\eqref{eq:lifetimeZeroB}. In the low temperature regime (blue), however, the operator lifetime becomes those found in \eqref{eq:lifetimeEparallel}.      
}
\label{fig:lifetime-Eparal}
\end{figure}

What does this result tell us about the lifetime of the electric flux operator? While the integral $\phi(r_h)$ can be a dimensionless function of $T$ and $\CB$, the renormalised electromagnetic coupling can be chosen in such a way that $e_r^{-2} \gg \phi(r_h)$ and $e_r^{-2} T^2/\CB \ll 1$ so that $\omega \tau_{E^\parallel}\sim \omega/T \ll 1$. The second limit is essential as the matching procedure assumes that $\omega/T \ll 1$ and the solution outside this regime has to be discarded. Taking these factors into account, one concludes that the temperature dependence of the electric flux is different from the high temperature $T/\sqrt{\CB} \gg 1$ limit where $\tau_E \sim 1/T$ ( see Fig \ref{fig:lifetime-Eparal}). Naively taking the limit $T\to 0$  in \eqref{eq:lifetimeEparallel} will result in the vanishing lifetime of the electric flux in contrast to the result in \eqref{eq:lifetimeZeroB}. However, one has to carefully remove the limit $\omega/T\ll 1$ in order to access the lower temperature limit $\omega/T \sim 1, \omega/\sqrt{\CB} \ll 1$. 
%

\subsubsection{Perturbation perpendicular to equilibrium magnetic field}
Unlike the previous case, the perturbation $\delta B_{xz}$ that corresponds to $E^\perp = \frac{1}{2}\epsilon^{yzx}\la J_{zx}\ra$ is coupled to the metric perturbation. This is manifest in the equations of motion 
\begin{equation}
    \begin{aligned}
\frac{d}{dr} \Big(r^2 f e^{-W} \delta B_{xz}' + \CB (\delta G^x_t)  \Big) + \frac{\omega^2 e^{-W}}{r^2f} \delta B_{xz} &= 0\,, \\
\frac{d}{dr} \left( e^{4V+W} (\delta {G^x_t})' + 4\CB \delta B_{xz} \right) &= 0\, ,
    \end{aligned}\label{eq:bulkeom-perpendicular}
\end{equation}
where $\delta G_{\mu\nu}$ denotes the metric perturbations. Note that the coupled perturbations $\{ \delta B_{xz},\delta G_{tx}\}$ and $\{B_{yz},\delta G_{yz} \}$ are equivalent due to $SO(2)$ symmetry in the plane perpendicular to the equilibrium magnetic field. Also, the second equation of motion in \eqref{eq:bulkeom-perpendicular} can be written in a total derivative form $d \pi_{tx}/dr = 0$ with $\pi_{tx}$ is related to the momentum $\la T^{tx}\ra$. Since we are working in the zero wavevector limit, the conservation of momentum implies that $\pi_{tx} = 0$ in Fourier space (which can be shown explicitly using the $rx-$component of the Einstein equation).

The solution for $\delta B_{xz}, \delta G_{tx}$ in the outer region can be found by using the property of the background geometry combined with the Wronskian method as in \cite{Grozdanov_2019}. To be more precise, one first notes that the time-independent solution of the magnetised black brane can be written in a total derivative form, which implies the existence of two radially conserved currents.  
\begin{subequations}\label{eq:bulkBGconserved}
\begin{align}
Q_1 = r^2 f(V'-W')e^{2V+W} + 2 \CB h(r)   &= 0\,,\\
Q_2 = e^{4V+W}\frac{d}{dr} \left(e^{-2V}r^2 f  \right) - 4\CB h(r)  &= sT \,, \label{eq:BGconserved-2}
\end{align}
where we write the equilibrium ansatz for the gauge field as $B = h(r)\, dt\wedge dz$ with gauge choice $h(r_h) = 0$, which, together with the horizon regularity, sets $Q_1 =0$. The relation between $h(r)$ and the 3-form field strength is 
\begin{equation}
    e^{2V-W} h' = \CB\,.
    \label{eq:BGconserved-3}
\end{equation}
\end{subequations}
More details on obtaining these radially conserved quantities can be found in e.g. \cite{Gubser:2009cg}. With this ansatz, we can compare  \eqref{eq:bulkeom-perpendicular} and \eqref{eq:bulkBGconserved} and find that one of the solutions of \eqref{eq:bulkeom-perpendicular} when $\omega/r\to 0$ are
\begin{equation}
    \delta B_{xz} = \Phi_1(r) = h(r) + \frac{sT}{4\CB} \, ,\qquad \delta G^x_t = \Psi_1(r) = -e^{-2V}r^2 f\, .
\end{equation}
One can use the Wronskian method to find find a pair of solution of \eqref{eq:bulkeom-perpendicular} that are linearly independent to $\{ \Phi_1,\Psi_1\}$. These solutions are 
\begin{align}\label{eq:Phi2andPsi2}
    \Phi_2(r) = \frac{1}{4\CB} - \int^{\infty}_r d\ttr \left( \frac{\CB e^{W(\ttr)} \Psi_2(\ttr)}{\ttr^2 f(\ttr)}  \right)\, , \qquad \Psi_2(r) = \Psi_1(r) \int^{\infty}_r d\ttr \left(\frac{e^{-W(\ttr)}}{\ttr^4f(\ttr)^2} \right)
\end{align}
As a result, the outer region solution can be written as 
\begin{equation}\label{eq:BandGsolOuter}
\begin{pmatrix}  
\delta B_{xz}^{outer} \\ (\delta G^x_t)^{outer} - \frac{1}{\CB}\CJ_{xz}  
\end{pmatrix} = 
c_1\begin{pmatrix} 
\Phi_1\\ \Psi_1 
\end{pmatrix} + c_2 \begin{pmatrix} \Phi_2\\ \Psi_2 \end{pmatrix}
\end{equation}
where $\CJ_{xz}:= (r^2 f e^{-W} \delta B_{xz}' + \CB \delta G^t_x)$ is an integration constant of \eqref{eq:bulkeom-perpendicular}  at $\omega =0$. One can substitute the $BTZ\times \mathbb{R}^2$ ansatz into the solution in \eqref{eq:BandGsolOuter} to check that $\Phi_1,\Psi_{1,2}$ are finite at $r=r_h$ while $\Phi_2$ is singular. It is convenient to separate out the singular part of $\Phi_2$ in the following form 
\begin{equation}
    \Phi_2(r) =  \phi_2(r)- \left( \frac{\CB e^W \Psi_2}{r^2f'}  \right)_{r=r_h} \log f(r) 
\end{equation}
where $\phi_2(r)$ is the integral in \eqref{eq:Phi2andPsi2} with the logarithmic divergence subtracted. 
The boundary condition where the source for both metric and 2-form gauge field fluctuation vanishes corresponds to the following values of $c_1$ and $c_2$
\begin{equation}
    c_1 = \frac{\CJ_{xz}}{\CB} \, ,\qquad c_2 = - 4\left(\frac{sT}{4\CB} + h(\Lambda) + \frac{\CB}{\hat\kappa(\Lambda)} \right) \CJ_{xz}
\end{equation}
One can also check that $\CJ_{xz}$ is identical to the one-point function $\la \delta J^{xz}\ra$ via the definition \eqref{eq:defSourceAndResponse}. Note also that the ratio $c_2/\CJ_{xz}$ is finite due to the cancellation of the logarithmic divergence of $1/\kappa(\Lambda)$ and that of the near boundary solution of $h(r)$, obtained via \eqref{eq:BGconserved-3}.

Let us also pointed out another way to organise the equations of motion for $\delta B_{xz}$. It turns out that \eqref{eq:bulkeom-perpendicular} can be combined into a single equation of motion that reduces to a total derivative form at $\omega =0$. Following the procedure in e.g. \cite{Davison:2015taa} and some manipulation, we find 
\begin{equation}
    \left( [e^{4V+W} \left(e^{-2V} r^2 f\right)']^2r^2 f e^{-W} \delta \tilde B_{xz}'  \right)' + \frac{\omega^2}{r^2 f e^W} [e^{4V+W} (e^{-2V}r^2 f)']^2 \delta \tilde B_{xz} = 0
    \label{eq:bulkeom-perpendicular-Decoup}
\end{equation}
where $\delta \tilde B_{xz} = \delta B_{xz}/[e^{4V+W} (e^{-2V} r^2 f)']$. The outer region solution of \eqref{eq:bulkeom-perpendicular-Decoup} is easily obtained and can be shown to be identical to those of \eqref{eq:BandGsolOuter}.

We can now proceed to the inner region solution. This can be found by solving Eq.\eqref{eq:bulkeom-perpendicular-Decoup} and one find 
\begin{equation}
    \delta B^{inner}_{xz} = c^H \exp\left(-\frac{i\omega}{4\pi T} \log f(r)  \right)\, .
\end{equation}
In the intermediate region, we apply the expansion in \eqref{eq:expandInnerToIntermediate}. The coefficients $c_1,c_2$ are related to $c^H$ via
\begin{equation}
\begin{aligned}
    \left(-\frac{i\omega}{4\pi T}  \right)c^H &= - \left( \frac{\CB e^W \Psi_2}{r^2 f'} \right)_{r=r_h} c_2\, ,\\
    c^H &= \left( \frac{sT}{4\CB} \right)c_1 +   \phi_2(r_h) c_2\, , 
\end{aligned}
\end{equation}
Substituting the form of $c_1,c_2$ in terms of $\la \delta J_{xz}\ra$, we can write the relations in a form similar to $\la \delta J_{xy}\ra$, namely
\begin{equation}
    \left(-i\omega + \frac{1}{\tau_{E^\perp}} \right) \la \delta J_{xz}\ra =0\,.
\end{equation}
In the case of vanishing sources, we can write $\frac{c_2}{c_1}= -4\CB \left(\frac{sT}{4\CB} + h(\Lambda) + \frac{\CB}{\kappa(\Lambda)}  \right)$ and the relaxation time of the electric field perpendicular to the equilibrium magnetic field is 
\begin{equation}
    \tau_{E^\perp} =\frac{\sqrt{3}}{2\pi T \CB \Psi_2(r_h) } \left[\frac{sT}{4\CB} \frac{c_1}{c_2}+  \phi_2(r_h)  \right]
\end{equation}

In contrast to the result at $e_r^{-2}\gg 1$ and zero equilibrium magnetic field in \cite{Hofman_2018,Grozdanov_2019}, the lifetime at strong magnetic field $\CB/T^2$ has a very different form. To see this, it is useful to examined that the combinations that enter the lifetime as follows
\begin{equation}\label{eq:EperpLimit}
    \Psi_2(r_h)\propto \frac{1}{\CB T^2} \, , \qquad \phi_2(r_h) \propto \frac{1}{\CB} \,,\qquad \frac{c_1}{c_2} \propto \frac{1}{\CB^2}\,\quad \text{for large} 1/\kappa(\Lambda) 
\end{equation}
with proportionality constants given by some numbers of order $\CO(1)$. In the limit of large electromagnetic coupling $1/\kappa(\Lambda)\gg 1 $ and $\CB/T^2\gg 1$, we find that this gives a short lifetime of the form $\tau_{E^\perp} \propto T/\CB$. However, the location of this decaying mode $\omega = -i/\tau_{E^\perp}$ lies outside the hydrodynamic regime $\omega/T\ll1$. Thus, one conclude that there are no modes with long lifetime in this regime\footnote{Note also that, if one were to perform this analysis for a perturbation in the holographic dual to a theory with zero-form $U(1)$ at $T>0,\mu = 0$ (as in \cite{Kovtun:2003wp}, see also \cite{Grozdanov_2019}), one would find a spectrum of the form $\omega \sim T$. This solution is spurious as it lies outside the hydrodynamic regime $\omega/T\ll 1$ and, in fact, is not present in the genuine spectrum obtained numerically at finite $\omega/T$ \cite{Kovtun:2005ev}.  }. 

\subsection{Checking $T \gtrsim 0$ limit in $\omega/\sqrt{|\bf{B|}}\ll 1$ regime} \label{sec:matchingwB}

While the result in the previous section strongly indicated that the electric flux lifetime becomes very short at extremely low temperature, the simplicity of the holographic model also allows us to extend the analysis beyond the usual hydrodynamic $\omega/T \ll 1$ regime. We will first show that the zero temperature theory does not support the purely decaying mode of the form $\omega = -i/\tau$ in the small $\omega/\sqrt{| {\bf B}|}$ regime. Next, we further extend the regime of validity to that of $\omega/\sqrt{|{\bf B}|} \ll 1$ but for arbitrary $\omega/T$. The purpose of the latter is to show that $\tau_E \propto T/\sqrt{|{\bf B}|}$ without relying on the $\omega/T\ll 1$ limit. 

\subsubsection{Zero temperature} \label{sec:zeroTem}

A simple argument for the non-existence of such a slowly decaying mode, is the presence of Lorentz symmetry at zero temperature on the $AdS_3$ submanifold in the deep infrared. On the other hand, one can also show this, using matching methods similar to those in \cite{DHoker:2010xwl,DHoker:2010onp,Davison:2013bxa}.      

\begin{figure}[tbh]
\centering
\includegraphics[width=0.75\textwidth]{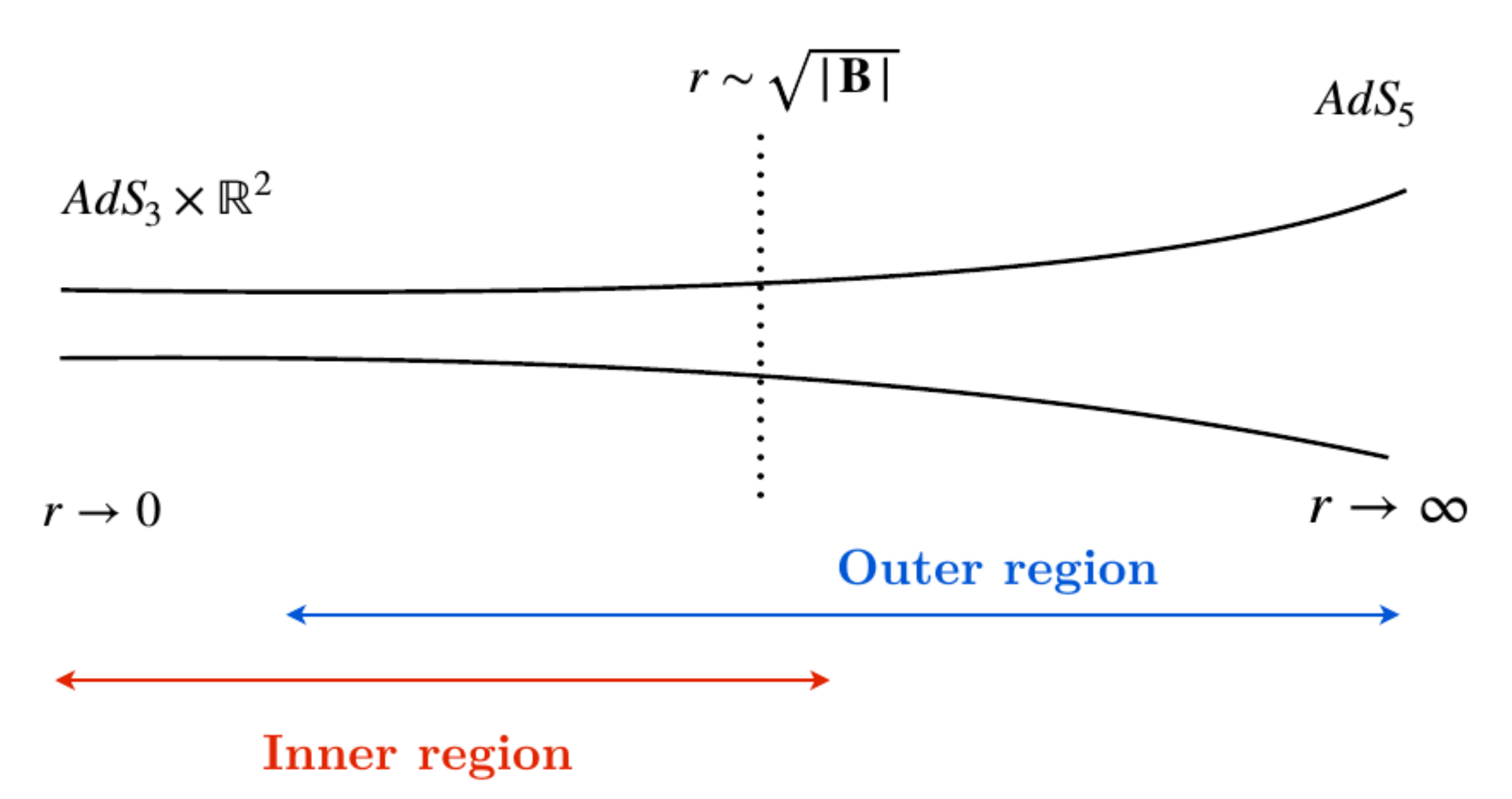}
    \captionsetup{justification=raggedright,singlelinecheck=false}
\caption{ 
A sketch of the bulk geometry at zero temperature. The inner region, whose solutions only depends on the ratio $\omega/r$ extended from the near horizon limit $r\to 0$ to the one where $\omega/r \sim \omega/\sqrt{|{\bf B}|} \to 0$ as we are working in the $\omega/\sqrt{|{\bf B}|}\ll 1$ limit. The outer region is defined to be the region where the $\omega^2/r^2$ and higher power in $\omega/r$ is suppressed, which can be extended toward $r\gg \sqrt{|\bf {B}|}$ as long as the frequency is small. 
}
\label{fig:bulkZeroT}
\end{figure}

To obtain this result, one first realises that the geometry of the magnetised black brane is that of an interpolation between IR $AdS_3\times \mathbb{R}^2$ and UV $AdS_5$. Roughly speaking, the IR geometry starts to becomes a good approximation as one starts to probe the scale below the magnetic field i.e. $r\sim \sqrt{|\bf {B}|}$. The inner and outer regions are defined such that they start off from the IR and UV geometry respectively, and extend to cover the overlap region (see Figure \ref{fig:bulkZeroT}). This is achievable when $\omega/\sqrt{|{\bf B}|}\ll1$.  

For concreteness, let us demonstrate how this works in the $E^{\parallel}$ channel that involves the bulk field $\delta B_{xy}$ governed by Eq.\eqref{eq:BxyBulkEom}. The solution can be written in the same form as \eqref{eq:outerBxy} evaluated at zero temperature (i.e. $r_h=0$). It is worth noting that the singular behaviour near $r/\sqrt{\CB} \to 0$ is different from that in earlier section. Instead, it can be written as 
\begin{equation}\label{eq:BxyOuterT0}
    \delta B_{xy}^{outer}(r) = c_1 - c_2\left( \log \Lambda - \bar\phi(r) + \frac{\CB/3}{6r^2}  \right) + \CO\left( \frac{\omega^2}{r^2}, \frac{\omega^2}{r^2} \log\left( \frac{\omega}{r}\right)\right)
\end{equation}
where the integration constants can be related to source and response via \eqref{eq:defSourceAndResponse}.
It is worth noting that the logarithmic divergence appears at order $\omega^2$. This is can be confirmed via Frobenius analysis in $AdS_5$ region (see e.g. \cite{Kovtun:2006pf}) and $AdS_3\times \mathbb{R}^2$ region (see appendix \ref{app:frobeniusAdS3R2}). The prefactor of the $r^{-2}$ divergence is obtained by evaluating $e^{2V(r)-W(r)}/f(r)$ at the horizon $r\to 0$. Here $\bar\phi(r)$ is the integral in \eqref{eq:outerBxy} subtracted by the $r^{-2}$ divergent and logarithmic divergent pieces. The resulting integral evalutated from $r=r_0\sim \sqrt{\CB}$ of the overlapping region to the UV cutoff $r =\Lambda$ is finite and its number is not extremely relevant for us as long as one keep $e_r^{-2}$ large. 

Next, we consider the inner region solution, which can be obtained by solving \eqref{eq:BxyBulkEom} in the $AdS_3\times \mathbb{R}^2$ region. Upon imposing horizon regularity at $r\to 0$, we find that the inner region solution is 
\begin{equation}
    \delta B_{xy}^{inner} = c^H \zeta K_1(\zeta) \, , \qquad \zeta = \frac{3\omega}{r}
\end{equation}

For these two branches of solutions to match, we extend the inner region solution to the regime where $\zeta = \omega/r \ll 1 $. We find that the `near boundary' expansion takes the form 
\begin{equation}
    \delta B_{xy}^{inner} = c^H \left( 1 + \frac{1}{2}\gamma \zeta^2 + \frac{1}{2} \zeta^2 \log \zeta + ...  \right)
\end{equation}
Matching this solution to the outer region, we find that $c_2\propto \omega^2$ unlike what happened in the previous section. Carrying on the matching procedure, we find that the polynomial governing the spectrum only depends on $\omega^2$ and thus rules out the purely imaginary mode $\omega = -i/\tau$. The same argument can also be made for the $E^\perp$ channel involving $\delta B_{xz}$. This is because, the part that is relevant to the matching procedure only depends on $\zeta^2$. See appendix \ref{app:frobeniusAdS3R2} for more details on the form of $\delta B_{xz}$ in the $AdS_3\times \mathbb{R}^2$ region.  

\subsubsection{$T \lesssim \omega \ll \sqrt{|{\bf B}|}$ limit }

\begin{figure}[tbh]
\centering
\includegraphics[width=0.75\textwidth]{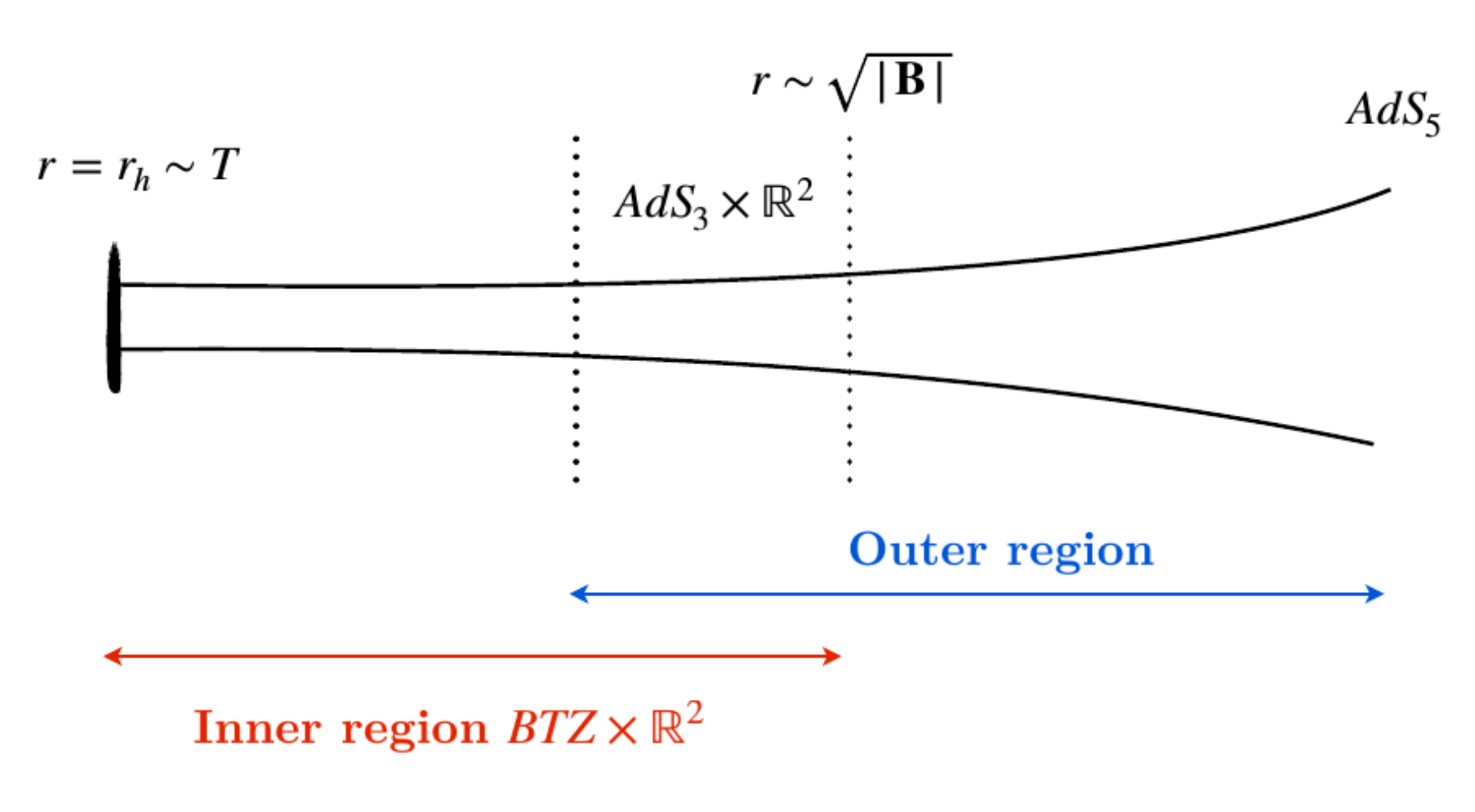}
    \captionsetup{justification=raggedright,singlelinecheck=false}
\caption{ 
A sketch of the bulk geometry at low temperature $T\ll 
\sqrt{|\bf{B}|}$. The inner region, whose solutions only depends on the ratio $\omega/r$ extends from the near horizon limit $r\to r_h 
\ll \sqrt{\mathbf{B}}$ to the one where $\omega/r \sim \omega/\sqrt{|{\bf B}|} \ll 1$, which corresponds to the near boundary region of $BTZ\times\mathbb{R}^2$ geometry, described by $AdS_3\times \mathbb{R}^2$. The outer region is defined to be the region where $\omega^2/r^2$ (and higher powers) is negligible and, therefore, can be extended toward $r\sim \sqrt{|\bf {B}|}$ in the $\omega/\sqrt{|\bf {B}|}\ll 1$ limit. 
}
\label{fig:bulkBTZ}
\end{figure}

In this section, we show that the electric flux lifetime can also be obtained regime where $\omega/T \gtrsim 1$ and $\omega/\sqrt{|\bf {B}|}\ll 1$ while keeping $\sqrt{|\bf{B}|}/T \gg 1$. The calculations closely resembles that of the zero temperature case except that the deep IR geometry is now $BTZ\times \mathbb{R}^2$ instead of $AdS_3\times\mathbb{R}^2$. Figure \ref{fig:bulkBTZ} illustrates this geometry where the $AdS_5$ joined with the $BTZ\times \mathbb{R}^2$ at the `boundary' $AdS_3\times \mathbb{R}^2$ of the IR geometry. We will only focus on the $E^\parallel$ fluctuations as it is the only channel that contains the decaying modes in the $\omega/T \ll 1$ regime. Similar computation for this type of geometry can also be found in \cite{DHoker:2010onp}.

The outer region solution, which extends from the UV $AdS_5$ to the intermediate $AdS_3\times \mathbb{R}^2$ region has the same form as  in \eqref{eq:BxyOuterT0}. This is possible only in the limit where $\sqrt{ \CB} \gg T$ so that $r/\sqrt{\CB}$ is always much greater than $T/\sqrt{\CB}\sim r_h/\sqrt{\CB}$ in this region. 

The inner region solution in the $BTZ\times \mathbb{R}^2$ region can be expressed in terms of a hypergeometric function (upon imposing ingoing boundary condition)
\begin{equation}
    \delta B_{xy}^{inner} = c^H \left( 1-\frac{r_h^2}{r^2} \right)^{-i\mathfrak{w}/2} \, {_2F_1}\left( -\frac{i\mathfrak{w}}{2},-\frac{i\mathfrak{w}}{2}, -\frac{i\mathfrak{w}}{2}; 1-\mathfrak{w}; 1-\frac{r_h^2}{r^2}\right)
\end{equation}
where $\mathfrak{w} = \omega/(2\pi T) = \omega/3r_h$. Extending this solution in the $r\gg r_h$ limit (which is possible due to $r_h/r \to 0$ as we approach the limit $\omega/r \to 0$) yields the following expansion \cite{abramowitz+stegun}
\begin{equation}
    \delta B_{xy}^{inner} \propto c^H \left[ 1+ \frac{i\omega r_h}{6 r^2} + \frac{1}{4} \left( \frac{\omega}{3r} \right)^2\left( 2-2\gamma -2 \psi(1-i\mathfrak{w}/2) - \log\left( \frac{r_h^2}{r^2} \right) \right) + \CO(\omega^3) \right]
\end{equation}
where $\psi(x)$ is the digamma function and the constants of proportionality are combinations of gamma functions that can be absorbed in the definition of $c^H$. The first two terms in $[...]$ are what important for us. By working to leading order in $\omega/r\ll 1$ as one approaches the intermediate $AdS_3\times \mathbb{R}^2$ region, we find the following matching solution
\begin{equation}
    c_1 -c_2 \log(\Lambda/\sqrt{\CB}) + \bar\phi = c^H \, , \qquad \left(\frac{\CB}{3}\right) c_2  = i\omega \left(\frac{2\pi T}{3}\right)c^H
\end{equation}
We can convert $c_1$ to the source $b_{xy}$ and $c_2$ as done in the previos sections. 
Upon taking $e_r^{-2}\gg \bar\phi$ (so that the solution lies in the regime of validity $\omega/\sqrt{\CB}\ll1$), we find the solution of the form $\omega = -i/\tau_{E^{\parallel}}$ where $\tau_{E^\parallel}$ is the same as in \eqref{eq:lifetimeEparallel}. This indicates that the lifetime indeed grows as $T/\sqrt{\CB}$ increases regardless of the ratio $\omega/T$.

\section{Conclusion}\label{sec:discuss}

The higher-form symmetry viewpoint of magnetohydrodynamics and its low temperature incarnation, the force-free electrodynamics, leads to new insights. The central focus of the present work was to established the absence of long-lived non-conserved operators. In turn, this indicates the validity of a hydrodynamic description at low temperature and strong magnetic field.  The question of whether the only operators that govern the deep IR dynamics are the conserved charges is important and ought to be asked before any quantitative attempt is made to study hydrodynamic properties (such as shear viscosity etc). All non-conserved operators must decay much faster than the scale of interest if a hydrodynamic interpretation is to be meaningful. 

We work with a holographic model which shares the same global symmetry as that of the plasma, namely only the energy, momentum and magnetic flux commute with the Hamiltonian. The model is simple enough for the lifetime of electric flux to be determined by classical bulk dynamics and the precise question is whether or not the electric flux is sufficiently long-lived to interfere with hydrodynamic modes.
Due to the anisotropy of the system in the presence of a strong expectation value of magnetic field, the lifetime of the electric field depends on its orientation. Our results can be summarised as follows
\begin{itemize}
    \item For electric flux $E_\parallel$ parallel to the magnetic field, the lifetime has a strong dependence on the double-trace coupling $\kappa$ which plays a role similar to the renormalised electromagnetic coupling. In the extreme limit of $e^{-2}_r \gg |\bf{B}|/T^2$, the lifetime can be large enough to be detectable by the analytic computation in both the `usual' hydrodynamic regime $\omega/T\ll 1$ and on even lower temperature regime where $\omega/\sqrt{\bf{B}}\ll1$ while $\omega/T$ may remains finite.
    We found that the lifetime becomes shorter as one decreases the ratio of $T/\sqrt{|\bf{B}|}$. The latter indicates that the lifetime will become extremely short in the extremely strong magnetic field regime $T/\sqrt{|\bf{B}|}\ll 1$ and cannot interfere with the low energy regime of $\omega/\sqrt{|\bf{B}|} \ll 1$ where the FFE limit is thought to be applicable. 
    \item For the component of electric flux $E_\perp$ perpendicular to the magnetic field, we find that there is no pole in the vicinity of $\omega/T \ll 1$. The dependence of the lifetime on the renormalised electromagnetic coupling disappears as one approaches the strong magnetic field limit. 
\end{itemize}
We also performed a consistency check at $T\to 0$ to ensure that there are no modes in the deep IR limit of $\omega/\sqrt{|\bf {B}|}\ll 1$. In this regime, the modes that indicate (potentially) long lifetime of $E_\parallel$ disappear from the low energy spectrum as anticipated. 

These computations are basic checks on the validity of FFE description. In the holographic context, it would be interesting to check if \textit{all} the accessible non-conserved operator truly have a parametrically short lifetime as well as confirming the low energy spectrum predicted by force-free electrodynamics (and its subsequent derivative corrections). Extraction of FFE effective action from gravity akin to \cite{Nickel:2010pr,Glorioso:2018mmw,deBoer:2018qqm} or the full constitutive relation as in \cite{Bhattacharyya:2008jc,Banerjee:2008th,Erdmenger:2008rm} would be desirable as a definitive proof of FFE description in the dynamically magnetised black brane geometry. Last but not least, it would be very interesting to investigate operators lifetime in (weakly coupled) quantum electrodynamics at finite $T$ and ${\bf B}$ to better understand FFE and its limitations in a system more directly connected to astrophysical plasma than the strongly coupled holographic model considered here.

\section*{Acknowledgements}

We would like to thank Jay Armas, Sa\v{s}o Grozdanov, Nabil Iqbal, Kieran Macfarlane, Watse Sybesma and L\'arus Thorlacius for helpful discussions and comments. We are particularly grateful to S. Grozdanov, N. Iqbal and L. Thorlacius for commenting on the manuscript. The work of N. P. was supported by Icelandic Research Fund grant 163422-052 and STFC grant number  ST/T000708/1. The work of A.R was supported in part by the Icelandic Research Fund under grant 195970-052 and by the University of Iceland Research Fund.

\begin{appendix}

\section{Numerical solution and evaluation of operators lifetime}\label{app:numerics}

In this section, remarks on the evaluation of the electric flux are elaborated. The numerical background solution for this geometry can be constructed in the same way as \cite{DHoker:2009mmn} using shooting method. The solution is a one-parameter family characterised by $\CB/T^2$ which allows us the freedom to choose $r_h=1, r_h^2f'(r_h)=1$ (or equivalently $T=1/4\pi$). It is also convenient to set $V(r_h)=W(r_h)=0$ which results in the UV boundary metric of the form 
\begin{equation}\label{eq:metricNumerics}
    \lim_{r\to \infty} ds^2 = r^2\left( -dt^2 + v(dx^2+dy^2) + w dz^2 \right) + \frac{dr^2}{r^2}
\end{equation}
Upon rescaling of spatial coordinates $\{dx,dy,dz\}\to \{dx/\sqrt{v},dy/\sqrt{v}, dz/\sqrt{w}\}$, we recover the desired background solutions. Note also that the physical magnetic flux is related to the input parameter (that produced the metric in \eqref{eq:metricNumerics}) by $\CB_\text{physical} = \CB_\text{input}/v$. 
A small caveat of this method is that one cannot find a smooth solution beyond $\CB_\text{input} \gtrsim \sqrt{3}/2$ which corresponds to the temperature $T/\sqrt{\CB} = (4\pi \sqrt{\CB_\text{input}/v})^{-1} \approx 0.05$. This is most likely an artifact of the presented numerical method as there exists a smooth solution in the zero temperature limit corresponding to the $AdS_3\times \mathbb{R}^2$ geometry in the deep IR. We should also note that this is a sufficiently low energy temperature as the entropy becomes sufficiently close to $s\propto T$ obtained from $BTZ\times \mathbb{R}^2$ geometry (c.f. \cite{DHoker:2009mmn,Grozdanov_2019b}). 
The background is generated for $r$ from $[1+10^{-3},10^6]$ and varying the (numerical) cutoffs within this order of magnitude does not change the obtained numerical results. 

Let us also remark on the the numerical value of the renormalised electromagnetic coupling $e_r^{-2} = \log(\Lambda/r_h) + \kappa(\Lambda)^{-1}$. This quantity strongly influences both the thermodynamics and low energy spectrum \cite{Grozdanov_2019b,Hofman_2018,Grozdanov_2019} of the model. In particular a small value of $e_r^{-2}$ would result in the speed of sound becoming imaginary \cite{Grozdanov_2019b}. Another way to see that this quantity should be large is to write it in terms of a renormalisation group independent scale $M_*$ that denotes the energy scale of a Landau pole \cite{Hofman_2018} i.e. $e_{r}^{-2} \sim \log(M_\star/T)$ where $M_\star \gg T$. We take this to be the largest scale in the problem--much larger than the accessible value of $\sqrt{\CB}/T$. 

\begin{figure}[tbh]
\centering
\includegraphics[width=0.7\textwidth]{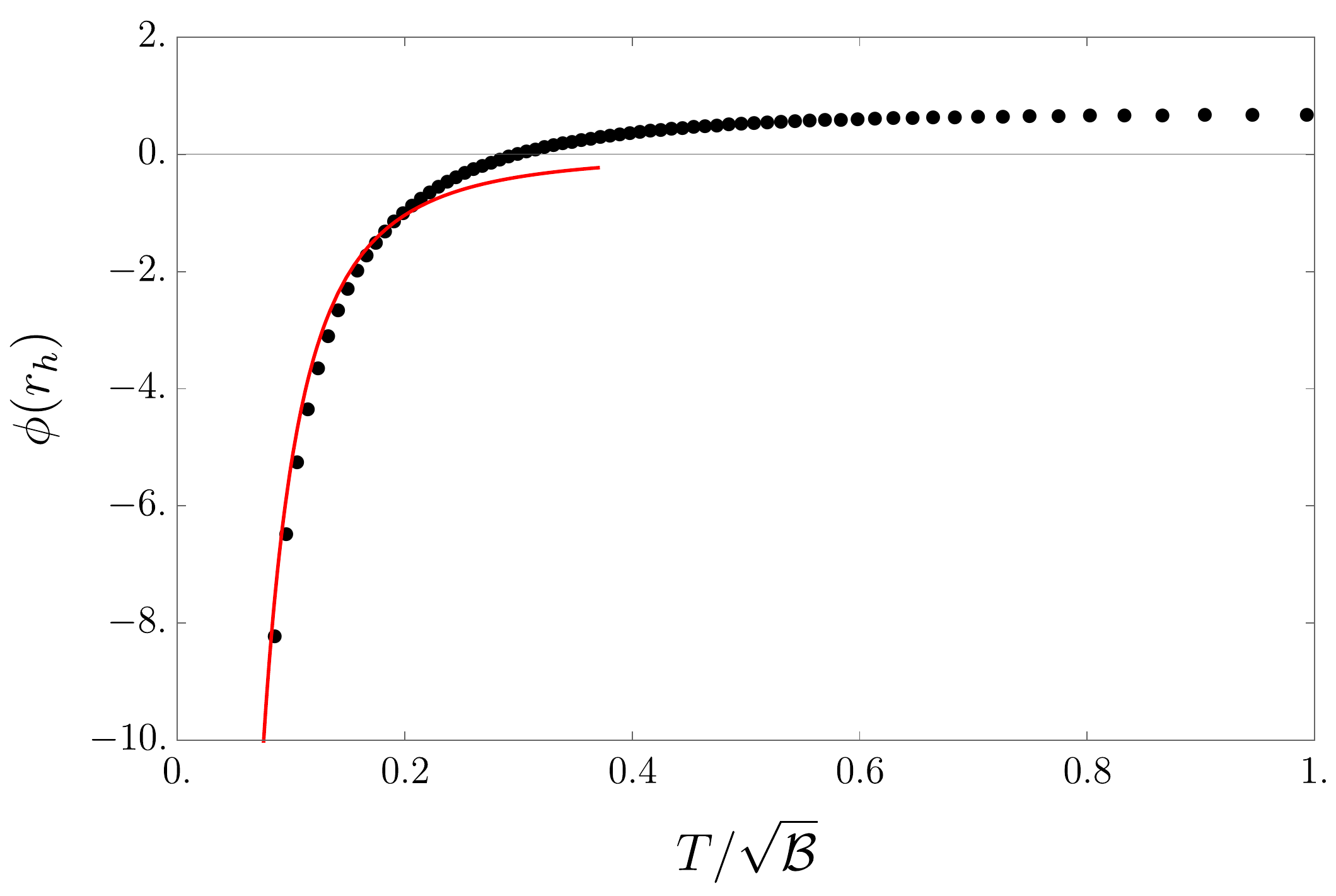}
    \captionsetup{justification=raggedright,singlelinecheck=false}
\caption{ Numerical evaluation of $\phi(r_h)$ in \eqref{eq:lifetimeEparallel} as a function of $T/\sqrt{\CB}$. The black dots denote the numerical evaluation while the red line denotes the fitting function for small $T/\sqrt{\CB}$ as $\phi \approx -(0.008)\frac{\CB}{T^2}\log(5.7 \CB/T^2) $. For high temperatures, the value of $\phi(r_h)$ is approximately constant around $0.69$. The value of $\phi(r_h)$ at lowest achievable temperature is at $\phi(r_h) =-23.49 $.  
}
\label{fig:PlotPhi}
\end{figure}

Numerical value of the integral for $\phi(r_h)$ in \eqref{eq:lifetimeEparallel} is shown in Figure \ref{fig:PlotPhi}. For a larger temperature (when $\phi(r_h) \approx \CO(1)$), the lifetime can be sensibly approximated  to be $\tau_{E^\parallel} \approx 2\pi (T/\CB) e_r^{-2}$. As $T/\sqrt{\CB}$ decreases, the lifetime becomes shorter and, if we are to extrapolate the fitting function $\phi \sim \frac{\CB}{T^2} \log\frac{\CB}{T^2}$ to even lower temperature where $e_r^{-2} \gtrsim \phi$, it will escape the regime of the validity of small $\omega/T,\omega/\sqrt{\CB}$ expansions. In this scenario, one shall conclude that there are no long-lived modes that can interfere with the low energy excitations.    
\section{Frobenius analysis in $AdS_3\times \mathbb{R}^2$ region}\label{app:frobeniusAdS3R2}

Consider the equation of motion for $\delta B_{xy}$ in the intermediate $AdS_3\times \mathbb{R}^2$ region:
\begin{equation}\label{eq:eomBxyAdS3R2}
    \delta B''_{xy}(r) + \frac{3}{r} \delta B_{xy}'(r) + \frac{\omega^2}{9r^2} \delta B_{xy}(r) = 0 
\end{equation}
The solution in this region can be obtained via Frobenius method. More precisely, one can change the radial coordinate into $\zeta= 3\omega/r$ and redefine $\delta B_{xy} = \zeta c(\zeta)$. It follows that $c(\zeta)$ is the solution of the Bessel equation of order $1$, which has a regular singular point at $\zeta =0$. The near-boundary $r\to\infty$, or equivalently $\zeta \to 0$, akin to the Fefferman-Graham expansion in the usual holographic renormalisation, can be written as 
\begin{subequations}

\begin{equation}\label{eq:solBxyAdS3R2}
\delta B_{xy}(\zeta) = c_1^M \CP_1(\zeta) +  \Big( c_2^M + \mathfrak{h} \log \zeta \Big)\CP_2(\zeta)
\end{equation}
where $c^M_1,c^M_2$ are integration constants and $\CP_i(\zeta)$ are regular polynomials of the  following form
\begin{equation}
    \CP_1 =  1 + \sum_{n=1}^{
    \infty} p_1^{[n]} \zeta^n \, , \qquad \CP_2 = \zeta^2\left( 1 +  \sum_{n=1}^\infty p_2^{[n]} \zeta^n \right)
\end{equation}
\end{subequations}

Similar to the usual procedure in the holographic renormalisation \cite{deHaro:2000vlm}, all the coefficients $p_1^{[n]}, p_2^{[n]}, \mathfrak{h}$ except $p_1^{[2]}$, which can be set to zero without loss of generality \cite{Kovtun:2006pf}, can be obtained recursively. The important piece of information here is the coefficient $\mathfrak{h}=1$ which can be obtained by recursively solving the equation \eqref{eq:eomBxyAdS3R2}. Another easy way to see this is to recast \eqref{eq:eomBxyAdS3R2} as the Bessel equation of order 1 as pointed out earlier. Then, using the fact that the Bessel functions $K_1(\zeta)$ and $I_1(\zeta)$ are two independent solutions of such equation and, for small $\zeta$ they admit the following asymptotic expansions (see e.g. \textsection 3.3 of \cite{bender2013advanced})
\begin{equation}
    I_1(\zeta) = \frac{\zeta}{2} + \frac{\zeta^2}{16} + \CO(\zeta)^3 \, , \qquad K_1(\zeta) = \left( \gamma + \log \frac{\zeta}{2} \right)I_1(\zeta)  +  \frac{1}{\zeta} 
\end{equation}
will result in the series expansions of the solution in $AdS_3\times \mathbb{R}^2$ region in \eqref{eq:solBxyAdS3R2}.

A similar procedure can also be applied for $E^\perp$ using Eq.\eqref{eq:bulkeom-perpendicular-Decoup}. Substituting $\delta\tilde B_{xz} = \zeta^2 c(\zeta)$, one finds that it obeys the Bessel equation of order $2$ whose $\zeta \ll 1$ expansion only yields even power in $\zeta$.

\end{appendix}

 \bibliographystyle{utphys}

\bibliography{biblio}

\providecommand{\href}[2]{#2}\begingroup\raggedright\begin{thebibliography}{10}

\bibitem{LLfluid}
L.~D. Landau and E.~M. Lifshitz, {\em {Fluid Mechanics}}.
\newblock Butterworth-Heinemann, 2nd~ed., 1987.

\bibitem{zwanzig1995}
R.~W. Zwanzig, {\em Statistical mechanics of irreversibility}.
\newblock Lectures on Theoretical Physics Volume 3, 139 (Interscience, 1961),
  1961.

\bibitem{forster1995}
D.~Forster, {\em Hydrodynamic Fluctuations, Broken Symmetry, and Correlation
  Functions}.
\newblock Perseus Books, 1995.

\bibitem{Hartnoll:2012rj}
S.~A. Hartnoll and D.~M. Hofman, ``{Locally Critical Resistivities from Umklapp
  Scattering},'' \href{http://dx.doi.org/10.1103/PhysRevLett.108.241601}{{\em
  Phys. Rev. Lett.} {\bfseries 108} (2012) 241601},
  \href{http://arxiv.org/abs/1201.3917}{{\ttfamily arXiv:1201.3917 [hep-th]}}.

\bibitem{Gaiotto:2014kfa}
D.~Gaiotto, A.~Kapustin, N.~Seiberg, and B.~Willett, ``{Generalized Global
  Symmetries},'' \href{http://dx.doi.org/10.1007/JHEP02(2015)172}{{\em JHEP}
  {\bfseries 02} (2015) 172}, \href{http://arxiv.org/abs/1412.5148}{{\ttfamily
  arXiv:1412.5148 [hep-th]}}.

\bibitem{Grozdanov_2017}
S.~Grozdanov, D.~M. Hofman, and N.~Iqbal, ``{Generalized global symmetries and
  dissipative magnetohydrodynamics},''
  \href{http://dx.doi.org/10.1103/PhysRevD.95.096003}{{\em Phys. Rev. D}
  {\bfseries 95} no.~9, (2017) 096003},
  \href{http://arxiv.org/abs/1610.07392}{{\ttfamily arXiv:1610.07392
  [hep-th]}}.

\bibitem{Schubring:2014iwa}
D.~Schubring, ``{Dissipative String Fluids},''
  \href{http://dx.doi.org/10.1103/PhysRevD.91.043518}{{\em Phys. Rev. D}
  {\bfseries 91} no.~4, (2015) 043518},
  \href{http://arxiv.org/abs/1412.3135}{{\ttfamily arXiv:1412.3135 [hep-th]}}.

\bibitem{Hernandez:2017mch}
J.~Hernandez and P.~Kovtun, ``{Relativistic magnetohydrodynamics},''
  \href{http://dx.doi.org/10.1007/JHEP05(2017)001}{{\em JHEP} {\bfseries 05}
  (2017) 001}, \href{http://arxiv.org/abs/1703.08757}{{\ttfamily
  arXiv:1703.08757 [hep-th]}}.

\bibitem{Armas:2018atq}
J.~Armas and A.~Jain, ``{Magnetohydrodynamics as superfluidity},''
  \href{http://dx.doi.org/10.1103/PhysRevLett.122.141603}{{\em Phys. Rev.
  Lett.} {\bfseries 122} no.~14, (2019) 141603},
  \href{http://arxiv.org/abs/1808.01939}{{\ttfamily arXiv:1808.01939
  [hep-th]}}.

\bibitem{Armas:2018zbe}
J.~Armas and A.~Jain, ``{One-form superfluids \& magnetohydrodynamics},''
  \href{http://dx.doi.org/10.1007/JHEP01(2020)041}{{\em JHEP} {\bfseries 01}
  (2020) 041}, \href{http://arxiv.org/abs/1811.04913}{{\ttfamily
  arXiv:1811.04913 [hep-th]}}.

\bibitem{dixon1982special}
W.~G. Dixon, {\em Special relativity: the foundation of macroscopic physics}.
\newblock CUP Archive, 1982.

\bibitem{anile2005relativistic}
A.~M. Anile, {\em Relativistic fluids and magneto-fluids: With applications in
  astrophysics and plasma physics}.
\newblock Cambridge University Press, 2005.

\bibitem{komissarov1999}
S.~S. Komissarov, ``{A Godunov-type scheme for relativistic
  magnetohydrodynamics},''
  \href{http://dx.doi.org/10.1046/j.1365-8711.1999.02244.x}{{\em Monthly
  Notices of the Royal Astronomical Society} {\bfseries 303} no.~2, (02, 1999)
  343--366}.

\bibitem{Arnold_2000}
P.~B. Arnold, G.~D. Moore, and L.~G. Yaffe, ``{Transport coefficients in high
  temperature gauge theories. 1. Leading log results},''
  \href{http://dx.doi.org/10.1088/1126-6708/2000/11/001}{{\em JHEP} {\bfseries
  11} (2000) 001}, \href{http://arxiv.org/abs/hep-ph/0010177}{{\ttfamily
  arXiv:hep-ph/0010177}}.

\bibitem{Blandford:1977ds}
R.~Blandford and R.~Znajek, ``{Electromagnetic extractions of energy from Kerr
  black holes},'' \href{http://dx.doi.org/10.1093/mnras/179.3.433}{{\em Mon.
  Not. Roy. Astron. Soc.} {\bfseries 179} (1977) 433--456}.

\bibitem{Komissarov_2004}
S.~Komissarov, ``{Electrodynamics of black hole magnetospheres},''
  \href{http://dx.doi.org/10.1111/j.1365-2966.2004.07446.x}{{\em Mon. Not. Roy.
  Astron. Soc.} {\bfseries 350} (2004) 407},
  \href{http://arxiv.org/abs/astro-ph/0402403}{{\ttfamily
  arXiv:astro-ph/0402403}}.

\bibitem{1969ApJ...157..869G}
P.~{Goldreich} and W.~H. {Julian}, ``{Pulsar Electrodynamics},''
  \href{http://dx.doi.org/10.1086/150119}{{\em \apj} {\bfseries 157} (Aug.,
  1969) 869}.

\bibitem{Wiegelmann_2012}
T.~Wiegelmann and T.~Sakurai, ``{Solar Force-free Magnetic Fields},''
  \href{http://dx.doi.org/10.12942/lrsp-2012-5}{{\em Living Rev. Sol. Phys.}
  {\bfseries 9} (2012) 5}, \href{http://arxiv.org/abs/1208.4693}{{\ttfamily
  arXiv:1208.4693 [astro-ph.SR]}}.

\bibitem{Gralla_2014}
S.~E. Gralla and T.~Jacobson, ``{Spacetime approach to force-free
  magnetospheres},'' \href{http://dx.doi.org/10.1093/mnras/stu1690}{{\em Mon.
  Not. Roy. Astron. Soc.} {\bfseries 445} no.~3, (2014) 2500--2534},
  \href{http://arxiv.org/abs/1401.6159}{{\ttfamily arXiv:1401.6159
  [astro-ph.HE]}}.

\bibitem{Compere:2016xwa}
G.~Comp\`ere, S.~E. Gralla, and A.~Lupsasca, ``{Force-Free Foliations},''
  \href{http://dx.doi.org/10.1103/PhysRevD.94.124012}{{\em Phys. Rev. D}
  {\bfseries 94} no.~12, (2016) 124012},
  \href{http://arxiv.org/abs/1606.06727}{{\ttfamily arXiv:1606.06727
  [math-ph]}}.

\bibitem{Uchida:1997}
T.~Uchida, ``Theory of force-free electromagnetic fields. i. general theory,''
  \href{http://dx.doi.org/10.1103/PhysRevE.56.2181}{{\em Phys. Rev. E}
  {\bfseries 56} (Aug, 1997) 2181--2197}.
  \url{https://link.aps.org/doi/10.1103/PhysRevE.56.2181}.

\bibitem{Thompson:1998ss}
C.~Thompson and O.~Blaes, ``{Magnetohydrodynamics in the extreme relativistic
  limit},'' \href{http://dx.doi.org/10.1103/PhysRevD.57.3219}{{\em Phys. Rev.
  D} {\bfseries 57} (1998) 3219--3234}.

\bibitem{Gralla_2019}
S.~E. Gralla and N.~Iqbal, ``{Effective Field Theory of Force-Free
  Electrodynamics},'' \href{http://dx.doi.org/10.1103/PhysRevD.99.105004}{{\em
  Phys. Rev. D} {\bfseries 99} no.~10, (2019) 105004},
  \href{http://arxiv.org/abs/1811.07438}{{\ttfamily arXiv:1811.07438
  [hep-th]}}.

\bibitem{glorioso2018effective}
P.~Glorioso and D.~T. Son, ``{Effective field theory of magnetohydrodynamics
  from generalized global symmetries},''
  \href{http://arxiv.org/abs/1811.04879}{{\ttfamily arXiv:1811.04879
  [hep-th]}}.

\bibitem{Benenowski:2019ule}
B.~Benenowski and N.~Poovuttikul, ``{Classification of magnetohydrodynamic
  transport at strong magnetic field},''
  \href{http://arxiv.org/abs/1911.05554}{{\ttfamily arXiv:1911.05554
  [hep-th]}}.

\bibitem{Grozdanov_2019b}
S.~Grozdanov and N.~Poovuttikul, ``{Generalised global symmetries in
  holography: magnetohydrodynamic waves in a strongly interacting plasma},''
  \href{http://dx.doi.org/10.1007/JHEP04(2019)141}{{\em JHEP} {\bfseries 04}
  (2019) 141}, \href{http://arxiv.org/abs/1707.04182}{{\ttfamily
  arXiv:1707.04182 [hep-th]}}.

\bibitem{Hofman_2018}
D.~M. Hofman and N.~Iqbal, ``{Generalized global symmetries and holography},''
  \href{http://dx.doi.org/10.21468/SciPostPhys.4.1.005}{{\em SciPost Phys.}
  {\bfseries 4} no.~1, (2018) 005},
  \href{http://arxiv.org/abs/1707.08577}{{\ttfamily arXiv:1707.08577
  [hep-th]}}.

\bibitem{Son:2002sd}
D.~T. Son and A.~O. Starinets, ``{Minkowski space correlators in AdS / CFT
  correspondence: Recipe and applications},''
  \href{http://dx.doi.org/10.1088/1126-6708/2002/09/042}{{\em JHEP} {\bfseries
  09} (2002) 042}, \href{http://arxiv.org/abs/hep-th/0205051}{{\ttfamily
  arXiv:hep-th/0205051}}.

\bibitem{Kovtun:2003wp}
P.~Kovtun, D.~T. Son, and A.~O. Starinets, ``{Holography and hydrodynamics:
  Diffusion on stretched horizons},''
  \href{http://dx.doi.org/10.1088/1126-6708/2003/10/064}{{\em JHEP} {\bfseries
  10} (2003) 064}, \href{http://arxiv.org/abs/hep-th/0309213}{{\ttfamily
  arXiv:hep-th/0309213}}.

\bibitem{Fuini:2015hba}
J.~F. Fuini and L.~G. Yaffe, ``{Far-from-equilibrium dynamics of a strongly
  coupled non-Abelian plasma with non-zero charge density or external magnetic
  field},'' \href{http://dx.doi.org/10.1007/JHEP07(2015)116}{{\em JHEP}
  {\bfseries 07} (2015) 116}, \href{http://arxiv.org/abs/1503.07148}{{\ttfamily
  arXiv:1503.07148 [hep-th]}}.

\bibitem{Janiszewski:2015ura}
S.~Janiszewski and M.~Kaminski, ``{Quasinormal modes of magnetic and electric
  black branes versus far from equilibrium anisotropic fluids},''
  \href{http://dx.doi.org/10.1103/PhysRevD.93.025006}{{\em Phys. Rev. D}
  {\bfseries 93} no.~2, (2016) 025006},
  \href{http://arxiv.org/abs/1508.06993}{{\ttfamily arXiv:1508.06993
  [hep-th]}}.

\bibitem{DHoker:2009mmn}
E.~D'Hoker and P.~Kraus, ``{Magnetic Brane Solutions in AdS},''
  \href{http://dx.doi.org/10.1088/1126-6708/2009/10/088}{{\em JHEP} {\bfseries
  10} (2009) 088}, \href{http://arxiv.org/abs/0908.3875}{{\ttfamily
  arXiv:0908.3875 [hep-th]}}.

\bibitem{Grozdanov_2019}
S.~Grozdanov, A.~Lucas, and N.~Poovuttikul, ``{Holography and hydrodynamics
  with weakly broken symmetries},''
  \href{http://dx.doi.org/10.1103/PhysRevD.99.086012}{{\em Phys. Rev. D}
  {\bfseries 99} no.~8, (2019) 086012},
  \href{http://arxiv.org/abs/1810.10016}{{\ttfamily arXiv:1810.10016
  [hep-th]}}.

\bibitem{Davison:2014lua}
R.~A. Davison and B.~Gout\'eraux, ``{Momentum dissipation and effective
  theories of coherent and incoherent transport},''
  \href{http://dx.doi.org/10.1007/JHEP01(2015)039}{{\em JHEP} {\bfseries 01}
  (2015) 039}, \href{http://arxiv.org/abs/1411.1062}{{\ttfamily arXiv:1411.1062
  [hep-th]}}.

\bibitem{Chen:2017dsy}
C.-F. Chen and A.~Lucas, ``{Origin of the Drude peak and of zero sound in probe
  brane holography},''
  \href{http://dx.doi.org/10.1016/j.physletb.2017.10.023}{{\em Phys. Lett. B}
  {\bfseries 774} (2017) 569--574},
  \href{http://arxiv.org/abs/1709.01520}{{\ttfamily arXiv:1709.01520
  [hep-th]}}.

\bibitem{Davison:2018nxm}
R.~A. Davison, S.~A. Gentle, and B.~Gout\'eraux, ``{Impact of irrelevant
  deformations on thermodynamics and transport in holographic quantum critical
  states},'' \href{http://dx.doi.org/10.1103/PhysRevD.100.086020}{{\em Phys.
  Rev. D} {\bfseries 100} no.~8, (2019) 086020},
  \href{http://arxiv.org/abs/1812.11060}{{\ttfamily arXiv:1812.11060
  [hep-th]}}.

\bibitem{Witten:2001ua}
E.~Witten, ``{Multitrace operators, boundary conditions, and AdS / CFT
  correspondence},'' \href{http://arxiv.org/abs/hep-th/0112258}{{\ttfamily
  arXiv:hep-th/0112258}}.

\bibitem{Berkooz:2002ug}
M.~Berkooz, A.~Sever, and A.~Shomer, ``{'Double trace' deformations, boundary
  conditions and space-time singularities},''
  \href{http://dx.doi.org/10.1088/1126-6708/2002/05/034}{{\em JHEP} {\bfseries
  05} (2002) 034}, \href{http://arxiv.org/abs/hep-th/0112264}{{\ttfamily
  arXiv:hep-th/0112264}}.

\bibitem{Davison:2013bxa}
R.~A. Davison and A.~Parnachev, ``{Hydrodynamics of cold holographic matter},''
  \href{http://dx.doi.org/10.1007/JHEP06(2013)100}{{\em JHEP} {\bfseries 06}
  (2013) 100}, \href{http://arxiv.org/abs/1303.6334}{{\ttfamily arXiv:1303.6334
  [hep-th]}}.

\bibitem{Moitra:2020dal}
U.~Moitra, S.~K. Sake, and S.~P. Trivedi, ``{Near-Extremal Fluid Mechanics},''
  \href{http://arxiv.org/abs/2005.00016}{{\ttfamily arXiv:2005.00016
  [hep-th]}}.

\bibitem{Gubser:2009cg}
S.~S. Gubser and A.~Nellore, ``{Ground states of holographic
  superconductors},'' \href{http://dx.doi.org/10.1103/PhysRevD.80.105007}{{\em
  Phys. Rev. D} {\bfseries 80} (2009) 105007},
  \href{http://arxiv.org/abs/0908.1972}{{\ttfamily arXiv:0908.1972 [hep-th]}}.

\bibitem{Davison:2015taa}
R.~A. Davison, B.~Goutéraux, and S.~A. Hartnoll, ``{Incoherent transport in
  clean quantum critical metals},''
  \href{http://dx.doi.org/10.1007/JHEP10(2015)112}{{\em JHEP} {\bfseries 10}
  (2015) 112}, \href{http://arxiv.org/abs/1507.07137}{{\ttfamily
  arXiv:1507.07137 [hep-th]}}.

\bibitem{Kovtun:2005ev}
P.~K. Kovtun and A.~O. Starinets, ``{Quasinormal modes and holography},''
  \href{http://dx.doi.org/10.1103/PhysRevD.72.086009}{{\em Phys. Rev. D}
  {\bfseries 72} (2005) 086009},
  \href{http://arxiv.org/abs/hep-th/0506184}{{\ttfamily arXiv:hep-th/0506184}}.

\bibitem{DHoker:2010xwl}
E.~D'Hoker, P.~Kraus, and A.~Shah, ``{RG Flow of Magnetic Brane Correlators},''
  \href{http://dx.doi.org/10.1007/JHEP04(2011)039}{{\em JHEP} {\bfseries 04}
  (2011) 039}, \href{http://arxiv.org/abs/1012.5072}{{\ttfamily arXiv:1012.5072
  [hep-th]}}.

\bibitem{DHoker:2010onp}
E.~D'Hoker and P.~Kraus, ``{Magnetic Field Induced Quantum Criticality via new
  Asymptotically AdS$_{5}$ Solutions},''
  \href{http://dx.doi.org/10.1088/0264-9381/27/21/215022}{{\em Class. Quant.
  Grav.} {\bfseries 27} (2010) 215022},
  \href{http://arxiv.org/abs/1006.2573}{{\ttfamily arXiv:1006.2573 [hep-th]}}.

\bibitem{Kovtun:2006pf}
P.~Kovtun and A.~Starinets, ``{Thermal spectral functions of strongly coupled
  N=4 supersymmetric Yang-Mills theory},''
  \href{http://dx.doi.org/10.1103/PhysRevLett.96.131601}{{\em Phys. Rev. Lett.}
  {\bfseries 96} (2006) 131601},
  \href{http://arxiv.org/abs/hep-th/0602059}{{\ttfamily arXiv:hep-th/0602059}}.

\bibitem{abramowitz+stegun}
M.~Abramowitz and I.~A. Stegun, {\em Handbook of Mathematical Functions with
  Formulas, Graphs, and Mathematical Tables}.
\newblock Dover, New York, ninth dover printing, tenth gpo printing~ed., 1964.

\bibitem{Nickel:2010pr}
D.~Nickel and D.~T. Son, ``{Deconstructing holographic liquids},''
  \href{http://dx.doi.org/10.1088/1367-2630/13/7/075010}{{\em New J. Phys.}
  {\bfseries 13} (2011) 075010},
  \href{http://arxiv.org/abs/1009.3094}{{\ttfamily arXiv:1009.3094 [hep-th]}}.

\bibitem{Glorioso:2018mmw}
P.~Glorioso, M.~Crossley, and H.~Liu, ``{A prescription for holographic
  Schwinger-Keldysh contour in non-equilibrium systems},''
  \href{http://arxiv.org/abs/1812.08785}{{\ttfamily arXiv:1812.08785
  [hep-th]}}.

\bibitem{deBoer:2018qqm}
J.~de~Boer, M.~P. Heller, and N.~Pinzani-Fokeeva, ``{Holographic
  Schwinger-Keldysh effective field theories},''
  \href{http://dx.doi.org/10.1007/JHEP05(2019)188}{{\em JHEP} {\bfseries 05}
  (2019) 188}, \href{http://arxiv.org/abs/1812.06093}{{\ttfamily
  arXiv:1812.06093 [hep-th]}}.

\bibitem{Bhattacharyya:2008jc}
S.~Bhattacharyya, V.~E. Hubeny, S.~Minwalla, and M.~Rangamani, ``{Nonlinear
  Fluid Dynamics from Gravity},''
  \href{http://dx.doi.org/10.1088/1126-6708/2008/02/045}{{\em JHEP} {\bfseries
  02} (2008) 045}, \href{http://arxiv.org/abs/0712.2456}{{\ttfamily
  arXiv:0712.2456 [hep-th]}}.

\bibitem{Banerjee:2008th}
N.~Banerjee, J.~Bhattacharya, S.~Bhattacharyya, S.~Dutta, R.~Loganayagam, and
  P.~Surowka, ``{Hydrodynamics from charged black branes},''
  \href{http://dx.doi.org/10.1007/JHEP01(2011)094}{{\em JHEP} {\bfseries 01}
  (2011) 094}, \href{http://arxiv.org/abs/0809.2596}{{\ttfamily arXiv:0809.2596
  [hep-th]}}.

\bibitem{Erdmenger:2008rm}
J.~Erdmenger, M.~Haack, M.~Kaminski, and A.~Yarom, ``{Fluid dynamics of
  R-charged black holes},''
  \href{http://dx.doi.org/10.1088/1126-6708/2009/01/055}{{\em JHEP} {\bfseries
  01} (2009) 055}, \href{http://arxiv.org/abs/0809.2488}{{\ttfamily
  arXiv:0809.2488 [hep-th]}}.

\bibitem{deHaro:2000vlm}
S.~de~Haro, S.~N. Solodukhin, and K.~Skenderis, ``{Holographic reconstruction
  of space-time and renormalization in the AdS / CFT correspondence},''
  \href{http://dx.doi.org/10.1007/s002200100381}{{\em Commun. Math. Phys.}
  {\bfseries 217} (2001) 595--622},
  \href{http://arxiv.org/abs/hep-th/0002230}{{\ttfamily arXiv:hep-th/0002230}}.

\bibitem{bender2013advanced}
C.~Bender and S.~Orszag, {\em Advanced Mathematical Methods for Scientists and
  Engineers I: Asymptotic Methods and Perturbation Theory}.
\newblock Springer New York, 2013.

\end{thebibliography}\endgroup


\end{document}